\documentclass[10pt]{iopart}
\usepackage{setstack}
\usepackage{epsf}
\usepackage{bbold}

\newcommand{\dpq}{ \frac{d^4 {\bf p}}{(2 \pi)^4}}

\begin{document}
\title[Strange Baryons in a hot and dense medium within the NJL model]
{Strange baryons in a hot and dense medium within the Nambu-Jona-Lasinio model}
\author{F. Gastineau $\;$  and J. Aichelin }
\address{SUBATECH \\
Laboratoire de Physique Subatomique et des Technologies Associ\'ees\\
UMR Universit\'e de Nantes, IN2P3/CNRS, Ecole des Mines de Nantes\\ 
4, rue Alfred Kastler,
F-44070 Nantes Cedex 03, France.}
\begin{abstract}
Using an extended version of the Nambu-Jona-Lasinio model we build a simple 
description of the baryons as diquark-quark bound states. 
First, a description of the diquarks in a dense and hot medium is presented.
 Then, we introduce the formalism for the baryons 
 based on the Faddeev equation associated with the so-called `` static 
approximation'' 
which finally gives a Bethe-Salpeter equation in the diquarks-quarks channel. 
By identifying the baryons with the bound states, we can obtain a description
 of their properties. 
In particular, we obtain the right mass spectrum for the proton, $\Lambda$, 
$\Xi$, and $\Sigma$ at T=0 and $\mu=0$. 
We extend the formalism  to finite temperature and density to obtain a 
description of the mass change of these baryons in the medium. proc         
\end{abstract}
\maketitle
\section{Introduction} 
Knowing the properties of hadrons in a hadronic environment is presently one
of the challenging problems in heavy ion physics. 
Experimentally, the first
results have been published with clear evidence that the properties of mesons
change considerable when they are placed into a hadronic medium
\cite{QM01}.
 Theoretically, most of the efforts are concentrated around the mesons, where 
 a substantial change of their rest mass is predicted.
Baryons have attracted less attention.  
However,  the knowledge of their properties at finite
density and temperature is a prerequisite for interpreting the experimental 
results of heavy ion collisions at ultra-relativistic energies.
 There, it is assumed that a plasma of quarks and gluons is formed during the
 reaction. 
This plasma then expands and cools. When it passes the critical temperature 
for the chiral and deconfined phase transitions, it hadronizes into a colorless
 gas of hadrons. 
Since baryon masses change much more as a function of the density as compared
 to meson masses, 
 the thermodynamic properties of the hadron gas and  
especially the particle abundances at this transition   will be strongly 
influenced by the baryon properties in a medium. The Nambu-Jona-Lasinio(NJL) 
model offers the possibility to determine the properties of the baryons at
 finite temperature and density. In this model the baryons  are described as 
 diquark-quark bound states.

\section{The NJL model} 
The NJL model provides a simple implementation of dynamically broken chiral
 symmetry. It has been successfully used for the description of mesonic states
 at low-energy \cite{KWK} and some work has been done for baryons at zero 
temperature and density \cite{Ishii95,Buck,Tjon94}.
We use an SU(3) version of the NJL Lagrangian adding some extra terms for the
 qq channel interaction:
\begin{eqnarray}\label{LA}
{\mathcal{L}}_{\bar{q} q} &=& G_S[ (\bar{\psi} \lambda^F \psi)^2 + (\bar{\psi}
 i \gamma^5 \lambda^F\psi)^2] + G_V[ (\bar{\psi} \lambda^F \gamma_{\mu} \psi)^2
 + (\bar{\psi} i \gamma^5 \gamma_{\mu} \lambda^F\psi)^2] \nonumber \\
&+& G_D [det \bar{\psi}(1+\gamma^5)\psi + \bar{\psi}(1-\gamma^5)\psi] \nonumber
 \\
{\mathcal{L}}_{q q} &=& G_{DIQ}[ (\bar{\psi}^c i\gamma_5 \lambda^A_F \lambda^{A
^\prime}_C \psi)(\bar{\psi} i \gamma^5 \lambda^A_F \lambda^{A^\prime}_C \psi^c)
] 
\end{eqnarray}
where $A,A^\prime=2,5,7$ projects on the color and flavor $\bar{3}$ channel.

The ratio $G_S/G_{DIQ}$ is in principle fixed by the Fierz transformation, but 
we choose to leave it as a free parameter. The model needs to be regularized 
with a 3-momentum cut-off $\Lambda$. 
The parameters are fixed in order to give  correct value for the pion decay 
constant, quark condensate, mesons masses. In this work we have used the follow
ing parameters: $m_{0,q}=3.95$ MeV, $m_{0,s}=148$ MeV, $G_{S}/\Lambda^2=1.92$
 MeV$^{-2}$, $G_D/\Lambda^2=10.0$ MeV$^{-2}$,  $G_{V}/\Lambda^2=3.55$ MeV$^{-2}
$, $\Lambda=708.$ MeV, $G_{DIQ}/G_{S}=0.73$
\section{The diquarks} 
Before describing the baryonic states, we must study the behavior of diquarks.
 They are  (q-q) bound states, and therefore not observable in nature do their
 lack of color neutrality. 
First, we determine their mass in a hot and dense medium using the same
 formalism as for the mesons \cite{KWK,Reh96}, the Bethe-Salpeter equation. 
 In the qq channel it looks like :
\begin{equation}
{ \mathcal{T}}(q^2)  = {\mathcal{K}} + \frac{i}{2} \int \frac{d^4p}{(2\pi)^4}
 \left[ {\mathcal{K}}S_F(p+q)S_F(-q) {\mathcal{T}}(q^2) \right] 
\end{equation}
where ${\mathcal{K}}  =  \Omega^a 2i G_{DIQ} \bar{\Omega}^b $ ($a,b$ describe 
diquarks) with $\Omega^i  =  (\lambda^B_C \otimes \lambda^A_F \otimes \Gamma^i
 C)$. Here, we limit ourselves to the scalar channel ($\Gamma^i=i\gamma^5$).
The axial contribution to the baryon, which is not negligible as already shown 
in \cite{Buck}, will be treated later.

In the NJL model RPA and Bethe-Salpeter equation are identical and can be 
rewritten as a geometric series. Therefore one finally gets a simple form for
 the T-matrix,
where the singularities determine the bound states. 
In order to determine their mass one need to solve the gap equation:
 $det(1+\Pi(q^2) K)=0$ for $m_d^2 = q^2$ where $m_d$ is the diquark mass.
 We have  introduced here the polarization function defined by :
$
\Pi^{ij}(q^2) = -\int^{\Lambda} \frac{d^4 q}{(2 \pi)^4}
 \left [ \bar{\Omega}^i_{\alpha^\prime \beta^\prime}
 S_F^{\alpha^\prime \delta^\prime}(p+q) S_F^{\beta^\prime \gamma^\prime}(-p)
 \bar{\Omega}^j_{\delta^\prime \gamma^\prime} \right ]
$

 Previous calculations have been done in \cite{Ishii95} for $T=0$ and $\mu=0$.

To extend this equation to finite temperature and density, we used the
 imaginary time formalism as described in \cite{Kap}.
In order to build baryon states we  need the coupling constant between a 
diquark and two quarks. It can be defined by $g^{-2} \propto (\partial \Pi(k^2) / \partial k^2) \arrowvert_{k^2=m_d^2}$.

\section{Baryons as diquarks-quarks bound state} 
\begin{figure}
\begin{center}
\begin{minipage}[b]{.38\linewidth}
\leavevmode
\epsfxsize=7cm 
\epsfysize=2cm
\epsfbox{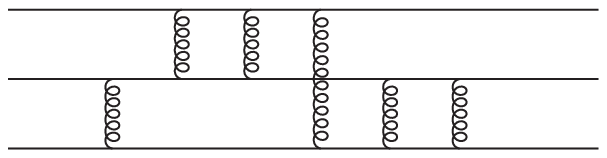}
\end{minipage}
\hfill
\begin{minipage}[b]{.38\linewidth}
\leavevmode
\epsfxsize=7cm 
\epsfysize=2cm
\epsfbox{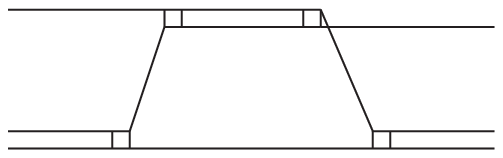}
\end{minipage}
\caption{Simple picture of the representation of baryons in our NJL model.
 Double lines are diquarks where as simple lines are quarks}
\label{fig:bar}
\end{center}
\end{figure}
The basic idea is to build baryons in the same way as  mesons or diquarks,
 using the Bethe-Salpeter equation.
To do this we need to know the diquark-quark interaction vertex which is not
 given in the Lagrangian. Therefore we have to add it in an empirical way.
We start from the basic  picture of a baryon (3 quarks interacting through
 gluon exchange) and apply it to the NJL model.
Each interaction is point-like, and  two interacting quarks form a bound state
 (Fig. \ref{fig:bar}). 
The result is  that quarks and diquarks interact by exchanging quarks.
Following ref. \cite{Buck} we assume  that the mass of the exchanged quark is
 infinite (``static approximation'').
 and therefore the interaction becomes point-like, with the vertex:
${\mathcal{Z}}^{\beta \gamma}_{bc}  =  g^{\prime}_{Dqq^\prime}\Omega^{\beta
 \delta}_a \left( -i/M_{ex} \right) \bar{\Omega}_b^{\gamma \alpha}
 g_{Dqq^{\prime}}$.
The Bethe-Salpeter equation for baryons reads:
\begin{equation}
{\mathcal{T}}_{ba}^{ \beta \alpha }({\bf P})  =  {\mathcal{Z}}_{ba}^{\beta \alpha} + 
 \int \frac{d^4 {\bf k}}{(2\pi)^4} {\mathcal{Z}}_{bc}^{\beta \gamma}
 S_{F l}^{\gamma \delta}(k)  S^l_{D,cd}(P-k) {\mathcal{T}}_{da}^{\delta \alpha}({\bf P}) ,
\end{equation}
where we have omitted the sum over the exchanged quarks, which is present if
 we study $\Lambda$(uds). 
Instead of a quark and an anti-quark propagator we have now the product of 
different propagators: one for a quark and one for a diquark. 
Following the usual approach to get the gap equation for baryons 
\begin{eqnarray}
det(1- \Pi Z)&=&0 \; \; \; \\
 { \rm where} \; \; \Pi(P^2) &=& -\int \dpq  T_i 
 \bar{\Omega}^{\alpha \delta}_c iS^{\delta \gamma}_{F}(p)iS_{D}^{cd}(P-p)
\Omega^{\gamma \beta}_d   T_j 
\end{eqnarray}
for $P^2 =m_B^2$. $ T_i,T_j$ are the isospin projections on the baryonic state.
The finite temperature and  density are taken into account  using the imaginary
 time formalism.

\section{Results}
The parameters of Lagrangian (\ref{LA}) lead to  a cross-over phase transition
 in the whole $(T,\mu)$ plane. 
Along the T-axis, $m_q(T)/ m_q(T=0)= 1/2$ for $T=220$ MeV. 
For the $\mu$-axis, $m_q(\mu)/ m_q(\mu=0)= 1/2$ for $\rho/\rho_0= 2.5$ and
 $T_c=22$0 MeV.

At $T= \mu = 0$ all the particles are stable with masses: $m_{q}=422 $MeV,
 $m_{s}=628$ MeV, $m_{qq}=557$ MeV, $m_{qs}=760$ MeV, $m_{nucleon}=938$ MeV,
 $m_{\Lambda}=1080$ MeV, $m_{\Sigma}=1181$ MeV and $m_{\Xi}=1278$ MeV.
\begin{figure}
\begin{center}
\epsfxsize=10cm  
\epsfysize=7cm
\epsfbox{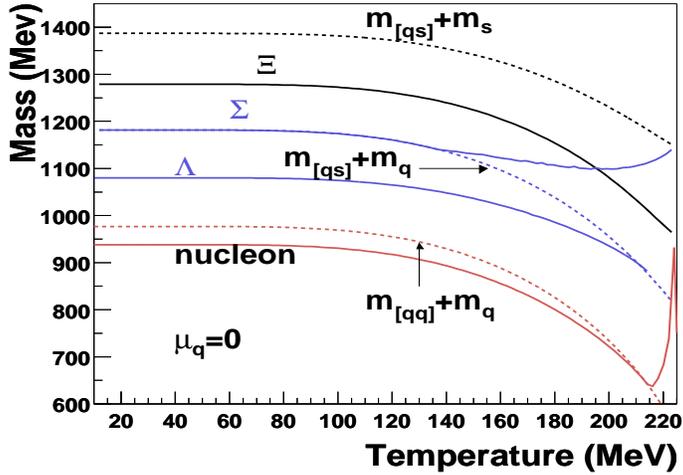}
\end{center}
\caption{Temperature dependence of the baryon masses (solid line). The dashed 
line represents the sum of the two baryon's constituent masses(diquark + quark).}
\label{fig:BarT}
\end{figure}

In Fig. \ref{fig:BarT}, we display the mass of baryons as a function of the temperature. 
All masses decrease with increasing temperature.
This mass change is due to  chiral symmetry restoration ($m_{q}$, $m_{s}$ go
 to their bare masses).
In contradiction to models which have baryons or mesons as degrees of freedom, 
the instability of baryons with respect to diquark-quark system (dashed line
 in Fig. \ref{fig:BarT}) can be calculated.
Nucleons and $\Lambda$ get unstable already at $T \approx 220$ MeV,  where as 
the $\Xi$ is stable longer due to its large strangeness content(s quark masses
 do not change too much with T). $\Sigma$ is a special case due to its weak 
binding energy, when the quark mass changes $\Sigma$ becomes unstable. 

\begin{figure}[hbt]
\begin{center}
\epsfxsize=10cm 
\epsfysize=7cm
\epsfbox{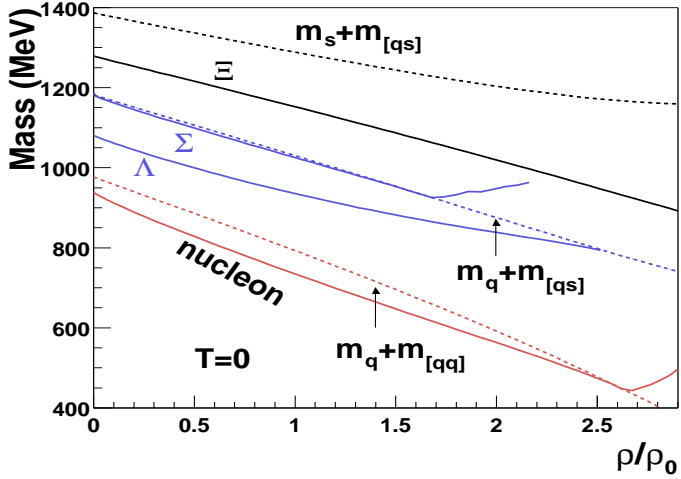}
\caption{Density dependence of the nucleon, $\Sigma, \Lambda, \Xi$ masses are
 shown(solid lines) and upper limit for the baryon masses(dashed lines): the
 sum of the quark and diquark masses for each particle.}
\label{fig:strange}
\end{center}
\end{figure}

In Fig. \ref{fig:strange} we plot the masses of baryons as function of the 
density for T=0.
As we observed in fig \ref{fig:BarT}, all  masses decrease and baryons become
 unstable for a given density ( $\rho / \rho_0 \approx 2.6$ for the nucleon
 ($M_{p,n} \approx 425$ MeV) and $\Lambda$ ($M_{\Lambda} \approx 850$ MeV); 
$\rho / \rho_0 \approx 1.7$ for $\Sigma$ ($M_{\Sigma} \approx 950$ MeV). 
The value of the nucleon mass at $\rho / \rho_0 = 1$ is about $700 MeV$. This 
is in agreement with the results of  Walecka or HQT models \cite{Serot} and
 also follow the results obtained in the other approaches \cite{Bentz,Zch} 
with the exception that in the NJL approaches we can also predict when baryons
 become unstable against description into a q - (qq) state.

\section{Conclusion} 
Employing the Nambu-Jona-Lasinio model we have constructed a simple formalism
 to described baryons as diquark-quark bound state. Extending the gap equation
 to finite temperature and density, we  show that baryon masses decrease by 
almost $30 \%$ close the chiral phase transition as compared to the vacuum 
values. This mass change may affect  baryon multiplicity when a plasma 
 hadronises into a hadron gas.
It can also modify the thermal freeze-out point as already shown in \cite{FlBr}. 
The next challenge is to describe the phase transition from a quark, diquark 
plasma into a hadron gas. A first step in this direction  has been done for 
the mesons \cite{Reh98}, but not yet for the baryons. Our simple model can 
easily be employed to study the phase transition at finite baryon density.

\end{document}